\documentclass[aip,graphicx]{revtex4-1}
\usepackage{graphics}
\usepackage{bm}
\usepackage{subfigure}
\usepackage{epsfig}
\usepackage[bookmarks,colorlinks,citecolor=blue,linkcolor=blue]{hyperref}

\usepackage{natbib}
\usepackage{siunitx}
\usepackage{soul}
\setstcolor{red}
\usepackage{array}
\usepackage{multirow}
\usepackage{amsmath}
\usepackage{esvect}
\usepackage{pdfcomment}
\usepackage{marginnote}
\usepackage{xcolor}
\definecolor{mygreen}{HTML}{2a9d8f}
\definecolor{myyellow}{HTML}{e09f3e}
\DeclareSIUnit\Molar{\textsc{m}}

\usepackage{xcolor}
\usepackage[normalem]{ulem}

\draft 

\begin{document}


\title{Robust Prediction of Frictional Contact Network in Near-Jamming Suspensions Employing Deep Graph Neural Networks} 



\author{Armin Aminimajd}
\affiliation{Department of Macromolecular Science and Engineering, Case Western Reserve University, Cleveland, OH, 10040, USA}
\author{Joao Maia}
\affiliation{Department of Macromolecular Science and Engineering, Case Western Reserve University, Cleveland, OH, 10040, USA}
\author{Abhinendra Singh}
\email{abhinendra.singh@case.edu}
\affiliation{Department of Macromolecular Science and Engineering, Case Western Reserve University, Cleveland, OH, 10040, USA}


\date{\today}

\begin{abstract}
The viscosity of the suspension consisting of fine particles dispersed in a Newtonian liquid diverges close to the jamming packing fraction. The contact microstructure in suspensions governs this macroscopic behavior in the vicinity of jamming through a frictional contact network (FCN). FCN is composed of mechanical load-bearing contacts that lead to the emergence of rigidity near the jamming transition. The stress transmission and network topology, in turn, depend sensitively on constraints on the relative motion of the particles. Despite their significance, predicting the FCN, especially close to jamming conditions, remains challenging due to experimental and computational impediments. 
This study introduces a cost-effective machine learning approach to predict the FCN using a graph neural network (GNN), which inherently captures hidden features and underlying patterns in dense suspension by mapping interparticle interactions. Employing a variation of GNN called the Deep Graph Convolutional Network (DeepGCN) trained on data-driven simulations, this study demonstrates robust generalization and extrapolation capabilities, accurately predicting FCNs in systems with divergent flow parameters and phase spaces, despite each being trained exclusively on a single condition. The study covers a wide range of phase space, from semi-dilute to jammed states, spanning transient to steady states, while systematically varying parameters such as shear stress ($\sigma_\text{xy}$), packing fraction ($\phi$) and sliding and rolling friction (\{$\mu_s, \mu_r$\}). The results of this research pave the way for innovative transferable techniques in predicting the properties of particulate systems, offering new avenues for advancement in material science and related fields.
\end{abstract}


\maketitle 

\section{Introduction}
Particulate suspensions, consisting of solid particles dispersed in a liquid medium, are ubiquitous in natural and industrial contexts. 
Examples range from everyday materials such as paints, concrete, food products, and pharmaceuticals to complex natural systems, including blood flow, mudflow, and lava~\cite{Coussot_1997, Jerolmack_2019, Galdi_2008, javid2024utilizing}.
During the last decade, a unifying framework has emerged, linking the physics of dense non-Brownian suspensions with dry granular rheology~\cite{Boyer_2011, guazzelli_2018, Morris_2020, Singh_2023}.
The key insight is that, under large deformations, the lubrication film between non-Brownian particles ruptures, leading to the formation of direct frictional contacts that share similarities with dry granular materials~\cite{Mari_2014, Singh_2018, Singh_2020, Morris_2020, Ness_2022, Guy_2015, Wyart_2014}.
Further studies have extended this understanding to the emergence of a frictional contact network (FCN) in dense suspensions, drawing parallels to classical two-dimensional granular material experiments by Behringer and coworkers~\cite{Bi_2011, Majmudar_2005,Majmudar_2007,Clark_2012,tordesillas2009buckling,daniels2008force,daniels2017photoelastic}.
\begin{figure*}
    \includegraphics[trim = 10mm 10mm 10mm 0mm, clip,width=1\textwidth,page=1]{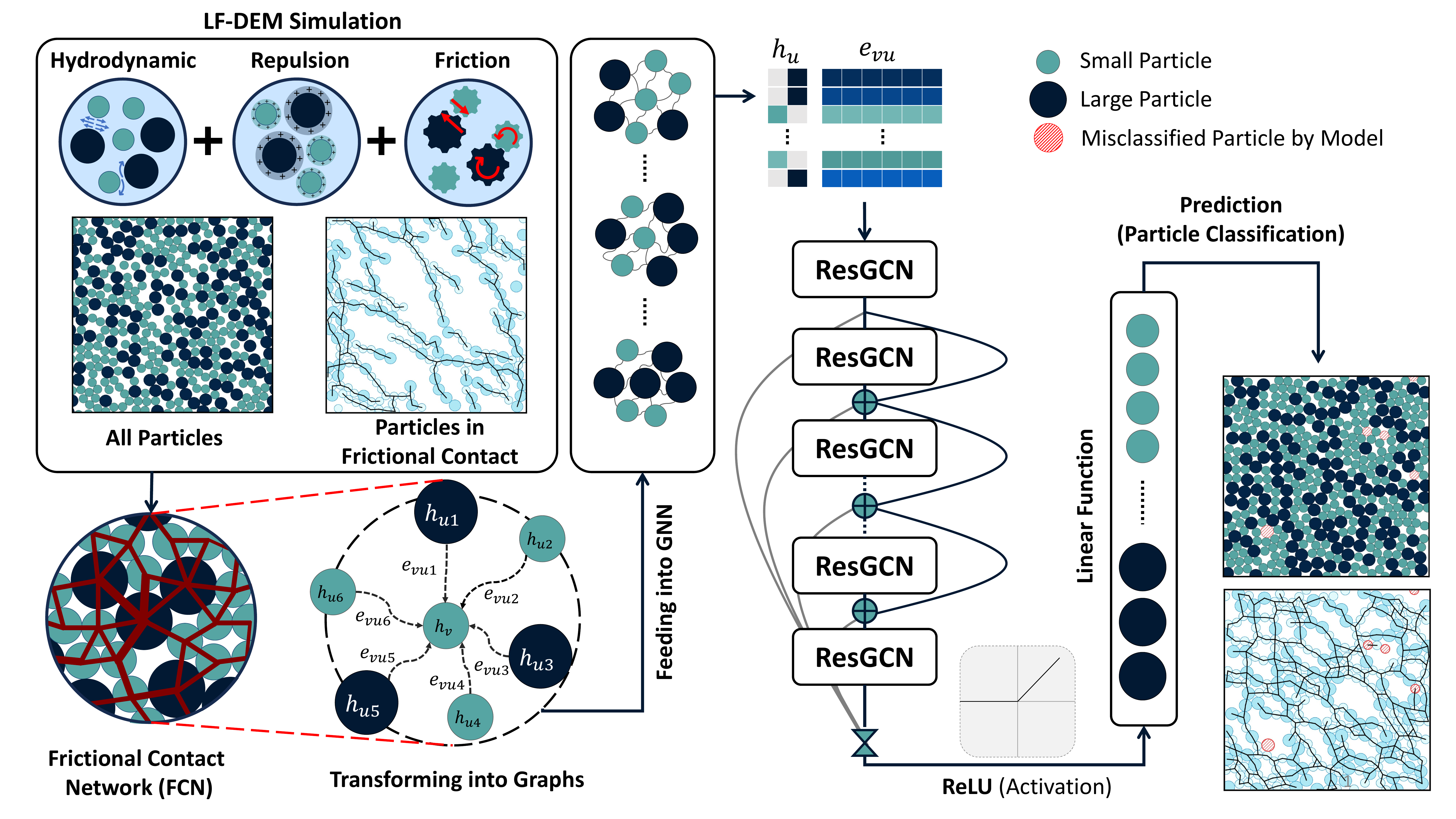}
    \caption{Schematic for the simulation process and the prediction of Frictional Contact Network (FCN) Using Deep Graph Convolutional Neural Network (DeepGCN). Initially, configurations are generated by the LF-DEM method, where particles obey Stokes flow (Eq.~\eqref{eq:1}). Then the configurations including the information of all particles (either in frictional contacts or not) are transformed into graphs readable for GNN where particles and the interparticle interactions are represented by nodes and edges, respectively. The graphs contain particle radius and relative distance between particles as node feature ($h_u$) and edge attributes ($e_vu$), respectively. Next, they are processed into residual graph convolution layers where nodes are updated through the message passing process using information of neighborhood nodes and edges. After normalization, a non-linear activation function (ReLU) applies to improve the model's understanding of complex underlying particle relationships. Then, a final linear layer classifies nodes to a specific class, whether a particle participates in the formation of FCN or not. Misclassified particles by the model, either incorrectly identified as part of the FCN or excluded when they should be included, are highlighted in red.} 
    \label{fig:fig1}
\end{figure*}

The suspension of interest in this study ranges from semi-dilute to dense systems near jamming. As the volume fraction $\phi$ increases, the viscosity increases, eventually diverging as volume fraction $\phi$ approaches the frictional jamming limit $\phi_J^\mu$ ~\cite{Peters_2016,Barik_2022, Singh_2019, ness2024dynamic}.
The frictional jamming point $\phi_J^\mu$ is influenced by various particle surface properties, including roughness, shape, interfacial chemistry, and particle size distribution, among  other factors~\cite{Morris_2020, Singh_2020, Singh_2022, Singh_2023, Singh_2024, James_2018, hsiao2019experimental, Hsu_2018}.
Recognizing the critical role of $\phi_J^\mu$ in dense suspension rheology, numerous experimental studies have explored ways to manipulate surface morphology~\cite{Hsiao_2017, hsiao2019experimental, Pradeep_2021, Hsu_2018} and interfacial chemistry~\cite{sivadasan2019shear, James_2018, James_2019, Chen_2022, Naald_2021, Xu_2020, lee2021microstructure}.
Fundamentally, these surface modifications alter the ``effective friction" by imposing constraints on the relative motion between particles, including translational, rotational, and twisting modes~\cite{Singh_2020, Guy_2018, Singh_2022, Singh_2023}. 

\begin{figure}
    \centering
    \includegraphics[trim = 0mm 300mm 340mm 0mm, clip,width=0.90\textwidth,page=2]{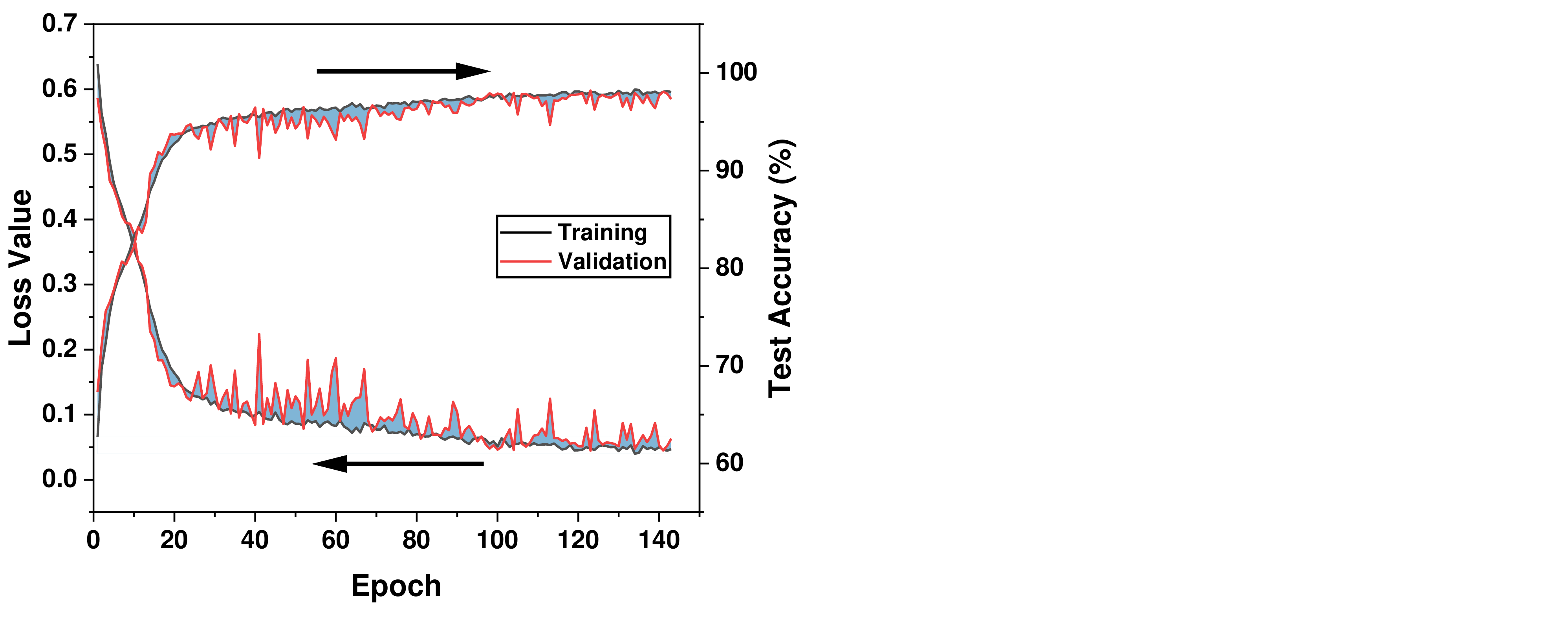}
    \caption{\textbf{Model Training Process for FCN Prediction.} An illustrative example of a model trained at $\phi=0.7$ and $\sigma/\sigma_0=5$, using a dataset of 320 graphs for training and 80 graphs for validation. As the loss value decreases over epochs, the prediction accuracy correspondingly improves. The training has been implemented using only two hidden layers with hidden dimensions set to 64. The process stops using the early stop technique that prevents overfitting.}
    \label{fig:fig2}
\end{figure}

The picture that emerges is that the constraints on relative motion stabilize the load-bearing force network, thereby enhancing the correlated motion~\cite{Singh_2020, LiuNagel_AnnRev}. In addition to that, these particle-level contacts give rise to large-scale force and contact networks that resist large deformations~\cite{Jamali_2020}.
Recently, network science tools have been employed to characterize the statistical properties of frictional force and contact networks and to correlate them with the rheological behavior of suspensions~\cite{Thomas_2018,thomas2020investigating,Nabizadeh_2022, Sedes_2022, Edens_2019, Edens_2021, Gameiro_2020, Naald_2024,Amico_2025}. 
Studies on dry granular materials and dense suspensions~\cite{Thomas_2018,thomas2020investigating,blumenfeld2003granular,blumenfeld2003granular} further suggest that the network space orthogonal to particle space may hold the key to establishing a statistical mechanics framework for out-of-equilibrium systems. 

Given this, significant efforts have been made to deduce interparticle forces utilizing both experimental and numerical approaches~\cite{Drescher_1972,howell1999stress, Majmudar_2005, hurley2016dynamic}. In dry granular materials, most studies employ the photoelastic technique to visualize the force chain network and measure the distribution of contact forces and particle stresses~\cite{Majmudar_2005}. However, only a few efforts have been able to extend to three-dimensional systems~\cite{Hurley_2014,li2024dynamic,lee2022force}. 
The application of these techniques to dense suspensions presents additional difficulties due to the small particle size $(R \le 1\mu)$m. The state-of-the-art techniques for generating 3D experimental contact networks involve using a photopolymer/initiator-doped suspension system to ``freeze'' the microstructure during shear~\cite{Pradeep_2021, Pradeep_2020, Nabizadeh_2024}. 
However, this method faces limitations, including competition between the photopolymerization protocol and suspension inertia when applied stress is removed. Furthermore, the cost and toxicity of the chemicals used in model suspension systems and photopolymers/initiators pose further challenges.
Beyond these experimental constraints, it remains difficult to precisely define direct frictional contacts between small non-Brownian particles - and consequently to predict the contact network -~\cite{Pradeep_2020}.
Although force and contact networks are more readily accessible in discrete particulate simulations, calculating interparticle interactions (such as lubrication, repulsion, and contact friction) requires highly efficient algorithms and parallel computing methods, particularly for larger system sizes close to jammed conditions. This makes simulations computationally expensive and time-consuming~\cite{yan2020scalable, monti2021fast}.

With the rapid advancement of data-driven approaches, there is a growing interest in applying these techniques in fluid mechanics, material science, and soft matter systems~\cite{ashwin2022deep, rossi2022identification, binel2023online,Jackson_2019,Barrat_2023,Ferguson_2017,Dulaney_2021,Gartner_2019,de2023data,floryan2022data,mahmoudabadbozchelou2021data,mahmoudabadbozchelou2024unbiased}.
Graph neural networks (GNNs), first introduced by Scarselli~\textit{et al.}~\cite{scarselli2008graph}, offer potential advantages over traditional techniques. It eliminates the need for handcrafted features by learning directly from the input microstructure and effectively modeling complex interactions and dependencies between particles, capturing both local and global structural information.
Then, the complex patterns and hidden features learned from the input data sets are utilized to predict unseen behavior or properties.
Despite their success in other soft matter systems, the use of GNNs to predict the properties of suspensions or granular materials is still emerging.
Mandal \textit{et al.} demonstrated the scalability of GNN in predicting the contact network in dry granular materials~\cite{Mandal_2022}. However, their model required extensive training and exhibited significant accuracy drops when extrapolating to unseen packing fractions or future strain conditions.
Similarly, Li  \textit{et al.} applied GNN for edge classification to estimate contact forces in 2D frictionless bidisperse granular materials at various pressure levels and volumetric mixing ratio~\cite{Li_2023_The_Prediction}. However, the highest accuracy they achieved (90\%) is restricted to low-pressure conditions.  
Recently, we introduced a machine learning framework that leverages the interconnected nature of graph structures to predict suspension properties~\cite{aminimajd2025scalability}. This GNN model demonstrated both scalability and robustness, achieving over 98\% accuracy in predicting frictional contact networks (FCNs) across different stress levels, particle sizes, and volumetric mixing ratios, even when trained under conditions significantly different from its initial training scenarios.
%
%
 
This study utilizes the proposed GNN framework to predict the frictional contact network (FCN) without requiring explicit knowledge of interparticle forces, as it only needs information about the relative distance of neighboring particles~\cite{aminimajd2025scalability}. 
Since the rheology and associated contact network are governed by externally applied shear stress and particle surface properties, we further demonstrate that the recently proposed GNN method can successfully extrapolate the FCN to unseen datasets, even in jammed conditions. 
This work provides a comprehensive analysis of FCN predictions across a broad spectrum of suspensions, ranging from semi-dilute to dense systems, by systematically varying key physical and processing parameters. These include the packing fraction ($ phi$), shear stress ($ sigma_textxy$), and constraints on relative motion, namely sliding $ mu_s$ and rolling $ mu_r$ frictions, all under steady-state conditions. We further show that for a given set of material parameters and the applied shear stress, the GNN model can predict future steady-state conditions--even training the machine exclusively on transient data.
To develop a mechanistic understanding of dense suspension flow, the rheological and microstructural results obtained from extensive simulations that span a wide parameter space are presented.
Unlike dry granular materials, two particles in contact (i.e., with finite non-zero overlap) in dense suspensions do not necessarily engage in frictional contact. This distinction is critical, as a normal force below a critical threshold normal force might still not activate the tangential frictional force~\cite{Mari_2014, Comtet_2017}.
The GNN model presented here is based on node classification to determine whether the two particles are in frictional contact. This classification enables the identification of particles that contribute to the formation of FCN without requiring explicit knowledge of the magnitude of the normal force or whether it exceeds the critical threshold or the critical force itself.
Additionally, this approach provides insight into the structure of the predicted FCN. 
The proposed method demonstrates a high degree of agreement with the FCN structures derived from simulation results across a wide range of configurations. Ultimately, the findings presented in this study offer a pathway to enhance the visualization of FCN, predict rheological properties, and analyze the microstructure of particulate systems. Finally, this approach addresses the challenges of traditional methods by achieving robust performance with limited training data, moderate computational resources, and reduced processing time. 

\section{Methods}

\paragraph*{{Simulating suspensions}:}
To simulate dense suspensions, the Lubrication Flow-Discrete Element Method
(LF-DEM), which integrates lubrication flow (LF) for hydrodynamic interactions with the discrete element (DEM) for contact force modeling, commonly used in dry granular materials, is employed~\cite{Seto_2013a, Mari_2014, Mari_2015, Morris_2020, Singh_2020}.
In our recent publications, we have demonstrated that the simulated suspensions in two- and three dimensions using this approach exhibit a comparable rheological response, provided the packing fraction $\phi$ is appropriately scaled relative to the jamming packing fraction $\phi_J$~\cite{aminimajd2025scalability,Gameiro_2020}.
To systematically explore the recently proposed GNN method~\cite{aminimajd2024scalability}, the simple-shear flow of non-Brownian frictional spheres immersed in a Newtonian fluid is simulated.
%
Given the Stokes flow regime, the particles obey the overdamped equation of motion:
\begin{equation}\label{eq:1}
    \vec{0} = \vec{F}_{\mathrm{H}}(\vec{X},\vec{U}) + \vec{F}_{\mathrm{C}}(\vec{X})
\end{equation}
where $\vec{X}$ and $\vec{U}$ denote the particle position and velocities, respectively. 
Here, $\vec{F}_{\mathrm{H}}$, $\vec{F}_{\mathrm{C}}$ correspond to the hydrodynamic and contact forces, respectively.
The hydrodynamic force calculations include both two-body lubrication forces and one-body Stokes drag. The leading term in calculating the lubrication forces diverges as $1/h$, where $h$ is the surface-to-surface distance between particles. Following previous studies~\cite{Melrose_1995, Morris_2020, Seto_2013a, Singh_2018, Singh_2020, Singh_2022}, it is assumed that lubrication breakdown occurs below $h_{\mathrm{min}}/a = 0.001$, where $a$ is the radius of the smaller particle. To account for this, lubrication forces are regularized below $h_{\mathrm{min}}$, allowing particles to form direct contacts.
The contact force is modeled using the traditional Cundall \& Strack approach \cite{Cundall_1979} and the algorithm by Luding~\cite{Luding_2008}. Specifically, contact forces are represented by linear springs and are activated only when $\delta^\text{(i,j)} = a_i + a_j - |r_i - r_j|>0$.

For a given particle pair ($i,j$) with overlap $\delta$ and a unit vector $\boldsymbol{n}$ connecting their centers, normal contact $\boldsymbol{F}_{\mathrm{C,N}}$, sliding friction force $\boldsymbol{F}_{\mathrm{C,T}}$, sliding friction torque $\boldsymbol{T}_{\mathrm{C,T}}$, and rolling friction torque $\boldsymbol{T}_{\mathrm{C,R}}$ are calculated using the following: 
\begin{subequations}
\begin{equation}
\boldsymbol{F}_{\mathrm{C,N}}^{(i,j)} = k_{n}\delta^{(i,j)}\boldsymbol{n}_{ij}~,
\end{equation}
\begin{equation}
\boldsymbol{F}_{\mathrm{C,T}}^{(i,j)} = k_{t} \boldsymbol{\xi}^{(i,j)}~,
\end{equation}
\begin{equation}
\boldsymbol{T}_{\mathrm{C,T}}^{(i,j)} = a_i \boldsymbol{n}_{ij} 
\times \boldsymbol{F}_{\mathrm{C,T}}^{(i,j)}~,
\end{equation}
\begin{equation}
\boldsymbol{T}_{\mathrm{C,R}}^{(i,j)} = 
a_{ij} \boldsymbol{n}_{ij} \times \boldsymbol{F}_{\mathrm{C,R}}^{(i,j)}~.
\end{equation}
\end{subequations}
%
%
The unit vector $\boldsymbol{n_{ij}} \equiv (\boldsymbol{r}_i - \boldsymbol{r}_j)/|\boldsymbol{r}_i - \boldsymbol{r}_j|$ points from the particle $j$ to $i$, with $a_{ij} \equiv 2 a_i a_j/(a_i + a_j)$ denoting the reduced radius.
$\boldsymbol{F}_{\mathrm{C,R}}^{(i,j)}= k_{r} \boldsymbol{\psi}^{(i,j)}$ 
is a quasi-force and is used exclusively for computing the torque $\boldsymbol{T}_{\mathrm{C,R}}^{(i,j)} $.
Here, $k_n$, $k_t$ and $k_r$ denote spring constants for  
the normal, sliding, and rolling, respectively.
Both sliding and rolling frictions obey Coulomb's friction laws, i.e., $|\boldsymbol{F}_{\mathrm{C,T}}^{(i,j)} \le \mu_s |\boldsymbol{F}_{\mathrm{C,N}}^{(i,j)}|$ and $|\boldsymbol{F}_{\mathrm{C,R}}^{(i,j)} \le \mu_r |\boldsymbol{F}_{\mathrm{C,R}}^{(i,j)}|$, where $\mu_s$ and $\mu_r$ are sliding and rolling friction coefficients, respectively.
Finally, the total contact force and torque are calculated as follows:
\begin{subequations}
\begin{equation}
\boldsymbol{F}_{\mathrm{C}}^{(i,j)} = 
\boldsymbol{F}_{\mathrm{C,nor}}^{(i,j)}  + \boldsymbol{F}_{\mathrm{C,slid}}^{(i,j)}~,
\end{equation}
\begin{equation}
\boldsymbol{T}_{\mathrm{C}}^{(i,j)} = 
a_i \boldsymbol{n}_{ij} \times \boldsymbol{F}_{\mathrm{C,slid}}
  + a_{ij}\boldsymbol{n}_{ij} \times \boldsymbol{F}_{\mathrm{C,roll}}~.
\end{equation}
\end{subequations}
Here $\boldsymbol{F}_{\mathrm{C,roll}}^{(i,j)}$ only contributes to the calculation of torque and hence the torque equation and does not contribute to the force equation.
The rate dependence is incorporated through the Critical Load Model (CLM), in which the normal force must exceed the critical force $F_0$ to activate interparticle frictions $\{\mu_s,\mu_r\}$. This critical force defines a characteristic stress scale, $\sigma_0 = F_0/a^2$which is used to nondimensionalize the shear stress.


The simulation consists of an assembly of $N=400$ non-Brownian bidisperse particles within a unit cell, employing Lees-Edwards periodic boundary conditions. The system comprises an equal volume of small particles (radius $a$) and large particles (radius $1.4a$), a choice that prevents crystallization while yielding rheological behavior consistent with experimental observations~\cite{Mari_2014, Singh_2020, Singh_2022}.
Simulations are performed under constant shear stress, resulting in a fluctuating shear rate $\dot{\gamma}(t)$. The relative viscosity in steady state is computed as $\eta_r(t) = \sigma/\eta_0\dot{\gamma}(t)$, with $\eta_0$ being the viscosity of the suspension liquid.

To train the model, the necessary dataset is generated by running simulations over a range of controlled parameters, specifically at fixed values of packing fraction ($\phi$), applied shear stress ($\sigma_\text{xy}$), sliding friction coefficient ($\mu_s$) and rolling friction coefficient $\mu_r$.
%

\paragraph*{{Machine Learning Method:}}
In this study, we employed a variant of the Graph Convolutional Neural Network (GCN) known as the Deep Graph Convolutional Neural Network (DeepGCN), introduced by Li \textit{et al.}~\cite{Guohao_2020, li_2019deepgcns}. This advanced architecture incorporates residual and dense connections, allowing deeper and more robust training of graph-based models compared to traditional algorithms.  
The schematic of the GNN training process has been depicted in Figure~\ref{fig:fig1}. To train the GNN model, after generation of the necessary configurations, using LF-DEM simulation the information of each configuration consisting of all particles in suspension is transformed into the graph, treating particles as nodes that include particle radii. We then draw edges ($e_\text{ij}$) between particles representing their interactions; these edges, termed edge attributes, include the distance between particles ($r_\text{ij}$), the x and y components of the vector $r_\text{ij}$, and the sine and cosine of the angle $r_\text{ij}$ forms with the flow direction.
Here, we focus only on the frictional contacts between the particles.
%
After converting datasets into graph representations, they are fed into the DeepGCN.
The DeepGCN model consists of $N_l$ layers with residual connections (ResGCN) incorporated. In each layer, the feature vector of a node is updated via a message-passing process that aggregates information from neighboring nodes and connected edges within the graph. The detailed equation of DeepGCN is provided in the following equations:

\begin{equation}
    h_v^{l} =  h_v^{(l-1)} + \sum_{u \in \mathcal{N}(v)} f(h_u^{(l-1)}, h_v^{(l-1)}, e_{uv})
    \label{eq:3}
\end{equation}

Here, $h_v^{(l)}$ is the updated node feature or hidden state for node $v$ at layer $l$, $ \mathcal{N}(v) $ denotes the neighborhood of node $v$, i.e., the set of nodes connected to $v$, $f$ is a function that takes as input the features of neighboring nodes i.e.,  $h_v^{(l-1)}$ and $h_u^{(l-1)}$ and their edge features $e_\text{uv}$, concatenating the node and edge features and applying non-linear aggregation function (Softmax). The output is then processed through a residual connection, adding the original node features $h_v ^l$ back to the result of the graph convolutional operation. This approach helps address vanishing gradient problems, facilitating the training of deep networks. 
Subsequently, after normalization, a nonlinear activation function, Relu, is applied to the output to introduce non-linearity, thereby enhancing the model's ability to learn. Finally, a linear function will be applied to the output to calculate the probability of each particle, indicating its participation in the FCN. This is a node classification task and the model output is one of the two possibilities, that is, the particle is in frictional contact (1) or not (0).
During training, we assessed the disparity between predicted and actual probability distributions of classes using a function called Binary Cross Entropy(BCE) (details are available \href{https://pytorch.org/docs/master/generated/torch.nn.BCEWithLogitsLoss.html\#torch.nn.BCEWithLogitsLoss}{here}). 
Figure \ref{fig:fig2} illustrates the training process as a function of epoch. The optimum goal is to minimize the loss value, reflecting efficient learning~\cite{mohammadagha2025machine}. Here, accuracy is defined as the total number of correctly classified particles (both true positives and true negatives~\cite{merikhipour2025transportation} i.e. correctly identify them whether they are in frictional contact or not) divided by the total number of particles in each configuration. However, the prediction results reported in the paper are the average prediction accuracy over the number of examples (configurations). To improve accuracy, Adam Optimizer with a learning rate of 0.005 is used to reduce the loss value in the training set. The early stop technique is used to stop the training process when the loss value does not reduce after 15 epochs by saving the last iteration parameters at which the loss is minimum. The training data set includes 320 graphs (80\% of the dataset), while a set of 80 configurations (20\% of the dataset) was utilized to validate the performance of the model. 
The model's hyperparameters are selected based on optimal performance on a validation set, comprising configurations not present in the training set. Subsequently, we assess the model's performance on a separate and independent test set beyond the initial training setting. \textcolor{black}{All the models were trained with only two layers, demonstrating that shallow message-passing with immediate neighbors is sufficient for individual particles to effectively capture the suspension’s hidden patterns.}
Further details about the details of machine learning technique and training process are documented in our previous study~\cite{aminimajd2024scalability}.

\begin{figure}
    \centering
    \includegraphics[trim = 0mm 120mm 480mm 0mm, clip,width=0.8\textwidth,page=3]{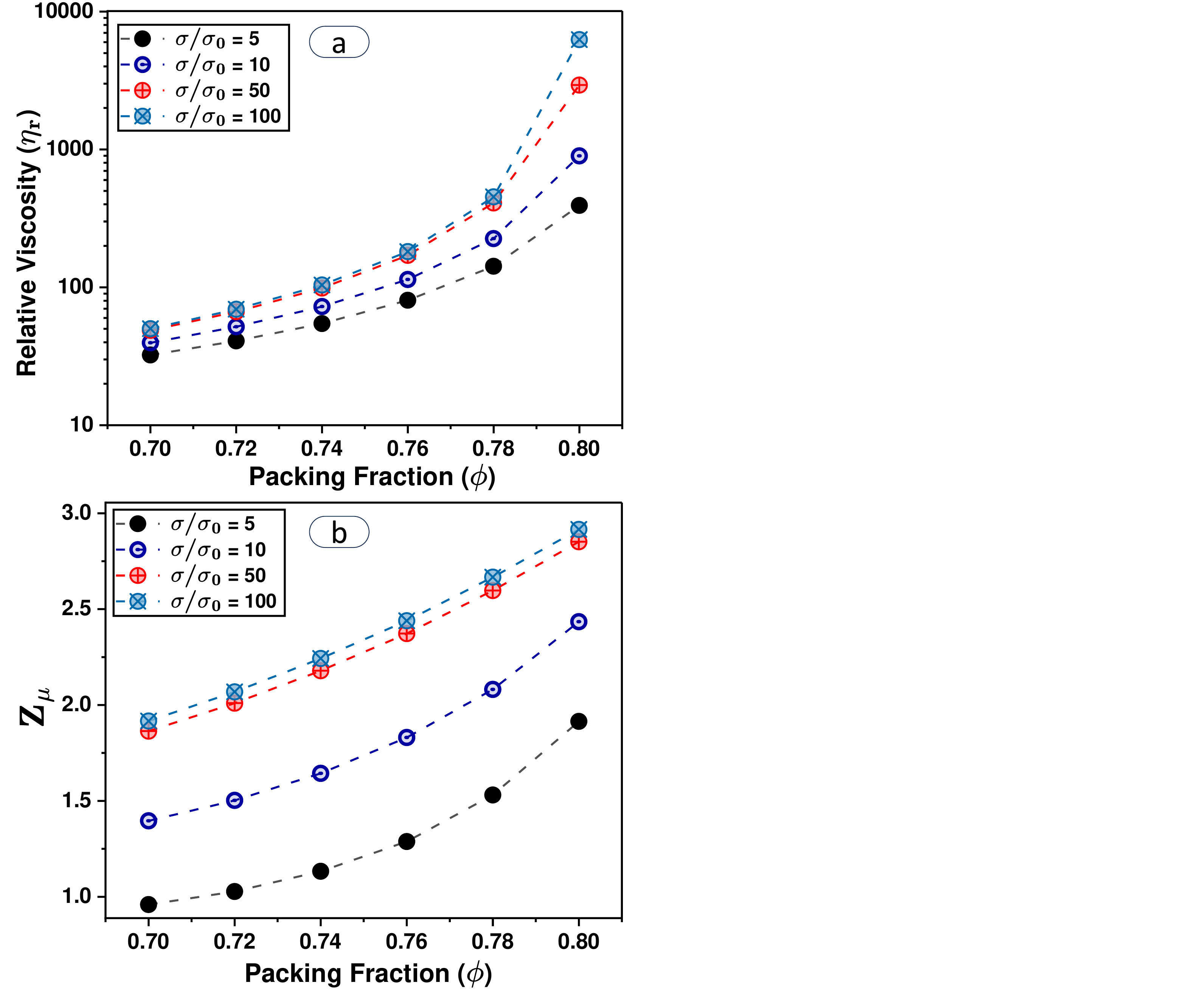}
    \caption{Rheological and microstructural behavior of suspensions as a function of packing fraction $\phi$ for different values of $\sigma / \sigma_0$. (a) relative viscosity $\eta_r$ and (b) average frictional coordination number $Z_\mu$. The simulations are performed for interparticle friction $\{\mu_s,\mu_r\} = \{0.5,0\}$.}
    \label{fig:fig3}
\end{figure}
\begin{figure*}
    \includegraphics[trim = 0mm 120mm 140mm 5mm, clip,width=1\textwidth,page=4]{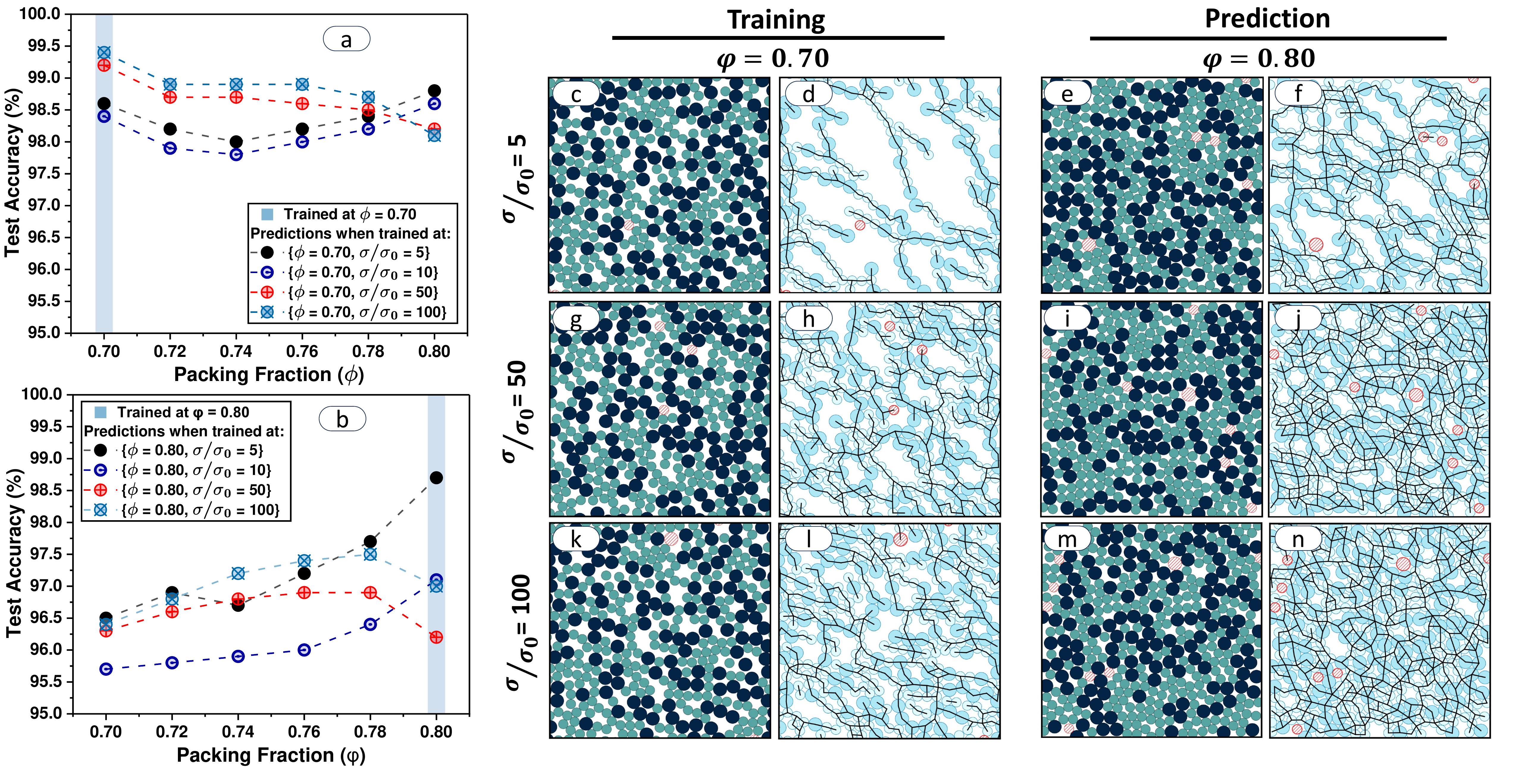}
    \caption{Test accuracy for the prediction of the frictional contact network (FCN) in suspensions across different packing fractions ($\phi$) at fixed shear stress values $\sigma / \sigma_0$. (a) Model trained at $\phi$ = 0.7 and evaluated for extrapolation up to $\phi$ = 0.8 at $\sigma / \sigma_0$ = 5, 10, 50, and 100. (b) similar to (a), but with the model trained at $\phi$ = 0.8 and evaluated for the extrapolation down to $\phi$ = 0.7. (c-n) display visualizations for the prediction of FCN with (c-d) the validation example for the model trained at $\phi = 0.7$ and fixed $\sigma / \sigma_0$ = 5, (e-f) and tested for the extrapolation of unseen configurations at $\phi = 0.8$ under the same  $\sigma / \sigma_0$. (c) is depicting all particles in suspension and (d) participating particles along their frictional contact network. Misclassified particles, either incorrectly identified as part of the FCN or excluded when they should be included, are highlighted in red. (g-j) and (k-n) the same as (c-f) but for $\sigma / \sigma_0$ = 50, and $\sigma / \sigma_0$ = 100, respectively. The simulations are performed for interparticle friction $\{\mu_s,\mu_r\} = \{0.5,0\}$.} 
    \label{fig:fig4}
\end{figure*}

\begin{figure}
    \centering
    \includegraphics[trim = 0mm 120mm 490mm 0mm, clip,width=0.8\textwidth,page=5]{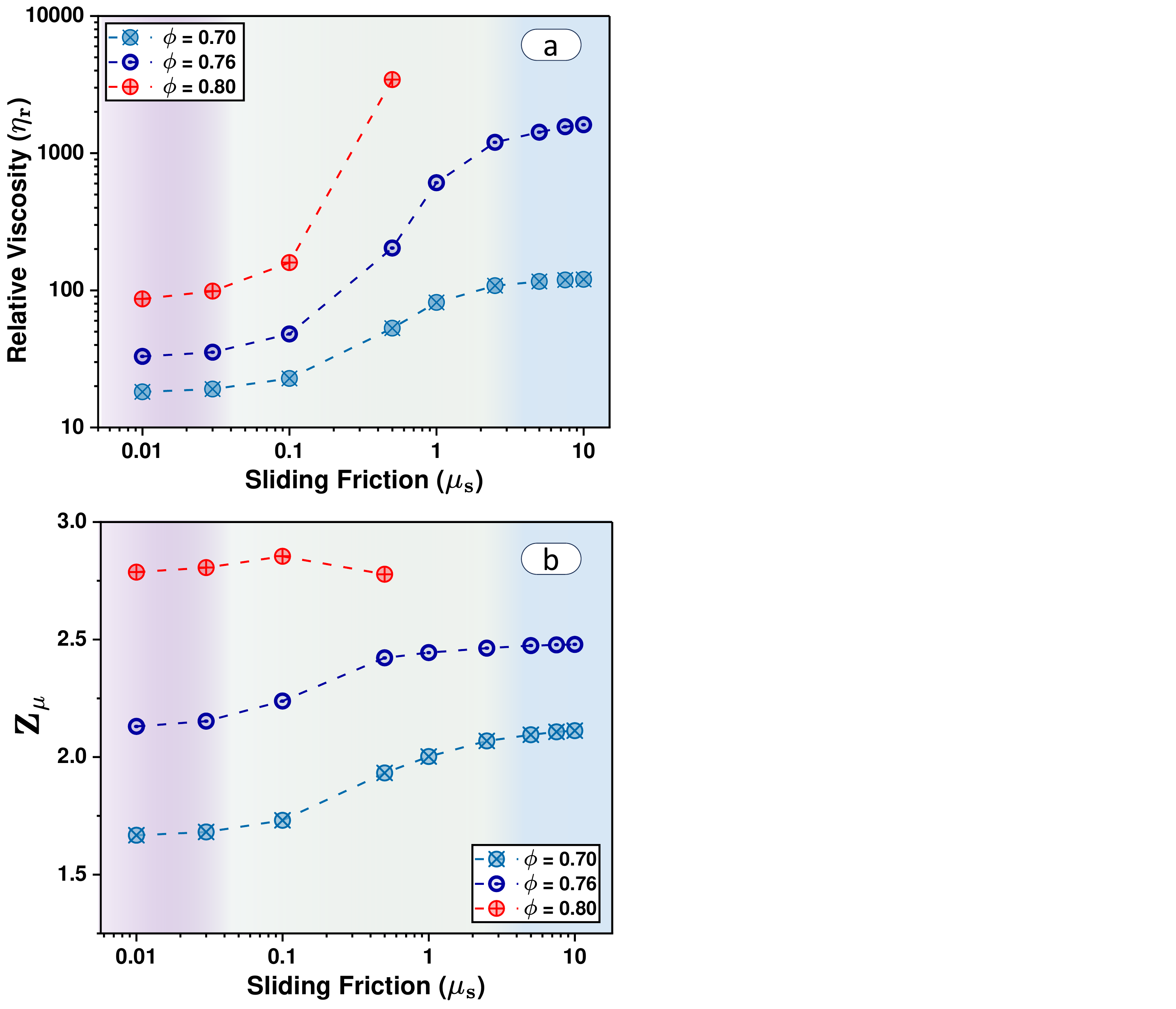}
    \caption{Rheological and structural behavior of suspensions from simulations as a function of sliding friction $\mu_s$ for different values of packing fractions $\phi$. (a) Relative viscosity $\eta_r$ and (b) average frictional coordination number $Z_\mu$. The color code reflect low, intermediate, and high friction regimes (from left to right).
    }
    \label{fig:fig5}
\end{figure}

\begin{figure*}
    \includegraphics[trim = 0mm 0mm 440mm 0mm, clip,width=0.85\textwidth,page=6]{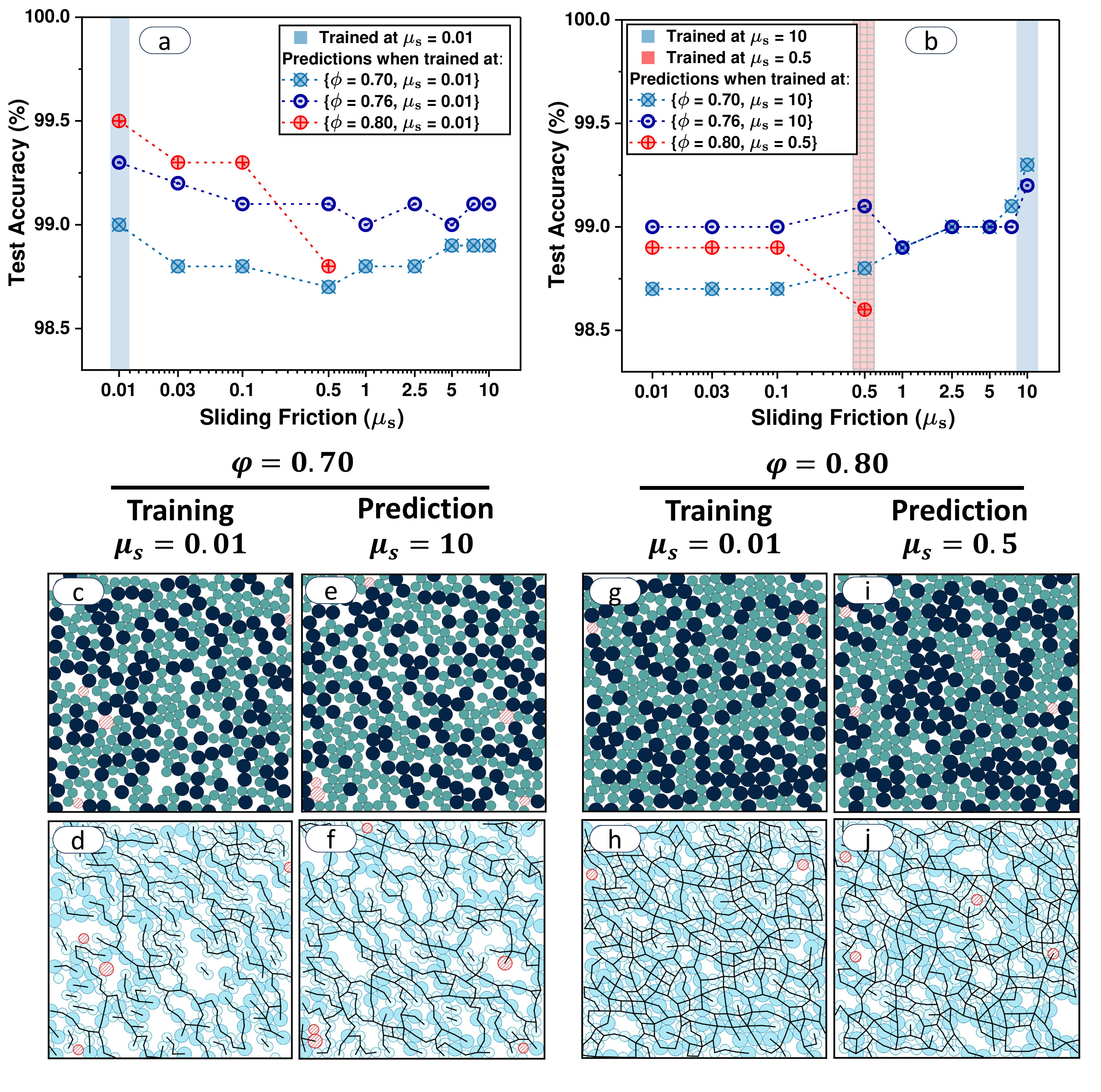}
    \caption{Test Accuracy for the prediction of frictional contact network (FCN) for suspensions at various values of sliding frictions ($\mu_s$) and fixed packing fractions ($\phi$). 
    (a) the models are trained at a fixed value of $\mu_s$ = 0.01 and $\phi$ = 0.7, 0.76, and 0.8 separately and evaluated for the extrapolation up to $\mu_s$ = 10, (b) the test accuracy results when the models are trained separately on $\phi$ = 0.7, 0.76 and $\mu_s$ = 10. Notably the suspension is shear-jammed at $\phi$ = 0.8 for $\mu_s \geq 0.5$  so the model for $\phi$ = 0.8 trained at $\mu_s$ = 0.5. (c-f) display visualizations for the prediction of FCN when the model is trained at fixed $\mu_s$ = 0.01 and $\phi$ = 0.7 (c-d) the validation and (e-f) tested for the extrapolation of unseen configurations at $\mu_s = 10$ under the same $\phi$. (c) is depicting all particles and (d) participating particles along their frictional contact network in suspension. (g-j) is the same as (c-f) but for when the model is conditioned at $\phi$ = 0.8.} 
    \label{fig:fig6}
\end{figure*}

\begin{figure}
    \centering
    \includegraphics[trim = 0mm 320mm 460mm 0mm, clip,width=0.8\textwidth,page=7]{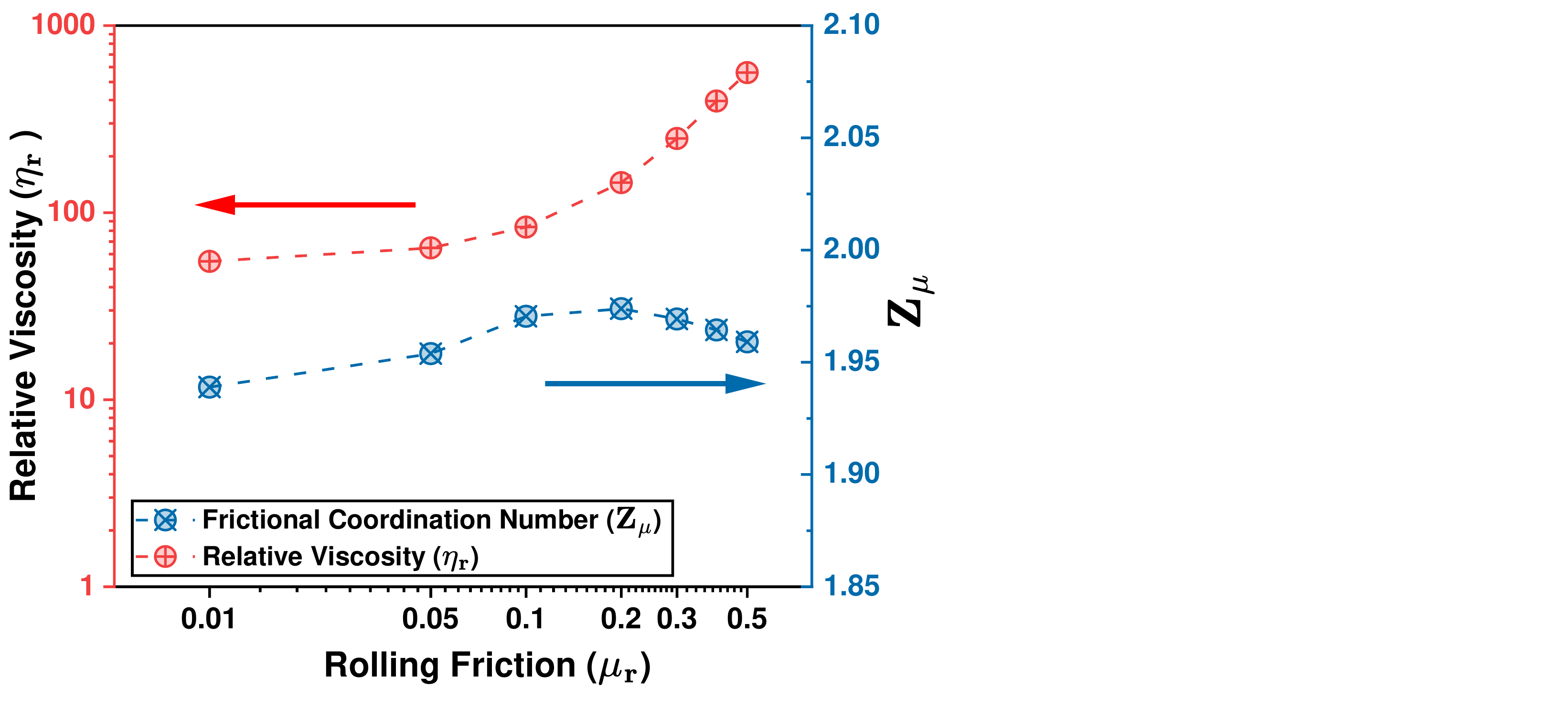}
    \caption{Relative viscosity $\eta_r$ (red) and average frictional coordination number $Z_\mu$ (blue) for suspensions as a function of rolling friction $\mu_r$ for packing fraction $\phi$ = 0.7 and $\mu_s=0.5$. Notably, the suspension is in a shear-jammed state for $\mu_r \geq$ 0.5.}
    \label{fig:fig7}
\end{figure}

\begin{figure*}
    \centering
    \includegraphics[trim = 0mm 250mm 250mm 0mm, clip,width=0.95\textwidth,page=8]{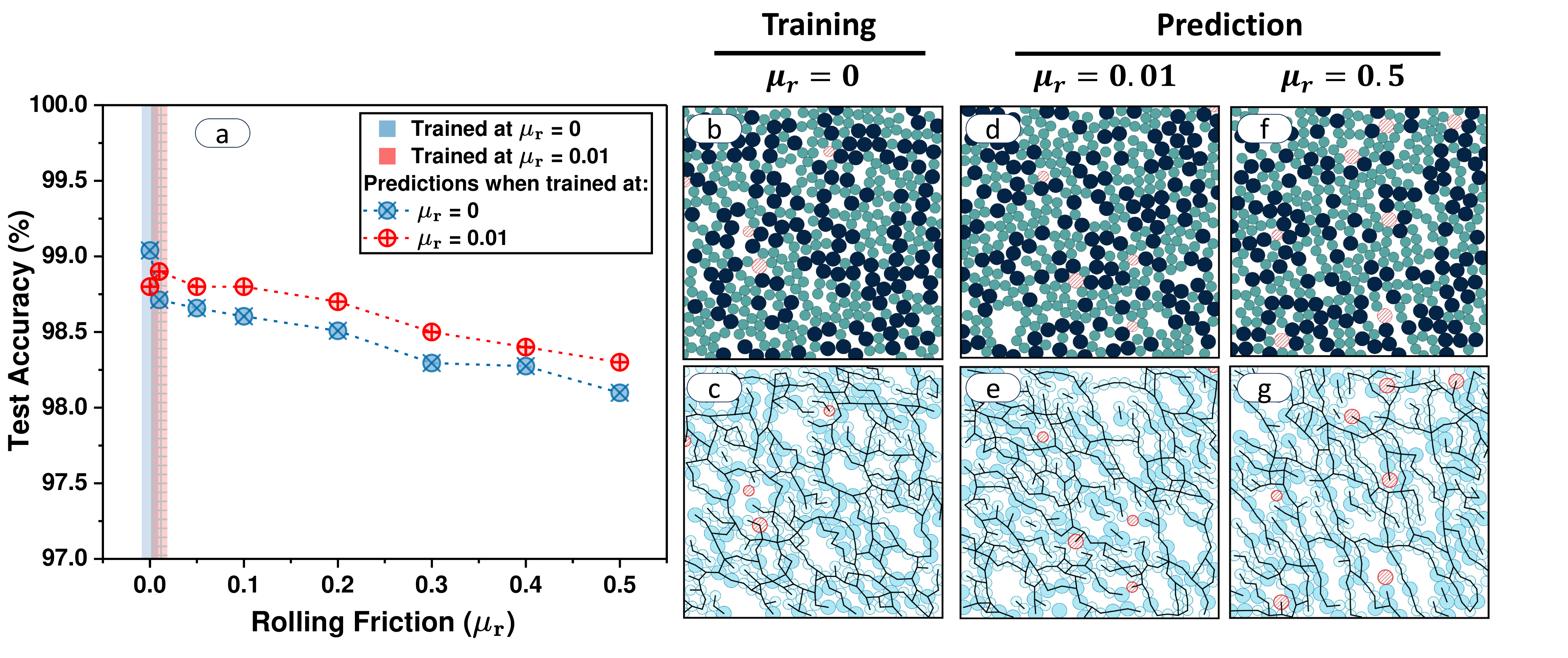}
    \caption{Prediction of frictional contact network (FCN) for suspensions at various values of rolling frictions ($\mu_r$) at fixed packing fraction $\phi=0.7$. (a) Test accuracy results when the model is trained at $\mu_r$ = 0 (blue) or $\mu_r$ = 0.01 (red). Notably the suspension is in a shear-jammed state for $\mu_r \geq$ 0.5. (b-g) Visualizations of predictions for the model trained at $\mu_r = 0$ and fixed $\phi$ = 0.70 with (b-c) validation at $\mu_r = 0$, (d-e) and (f-g) tested for the extrapolation of unseen configurations at $\mu_r$ = 0.01 and 0.5. (b, d, f) are depicting all the particles and (c, e, g)  participating particles along their frictional contact network in the suspension.} 
    \label{fig:fig8}
\end{figure*}

\begin{figure}
    \includegraphics[trim = 0mm 275mm 330mm 0mm, clip,width=0.85\textwidth,page=9]{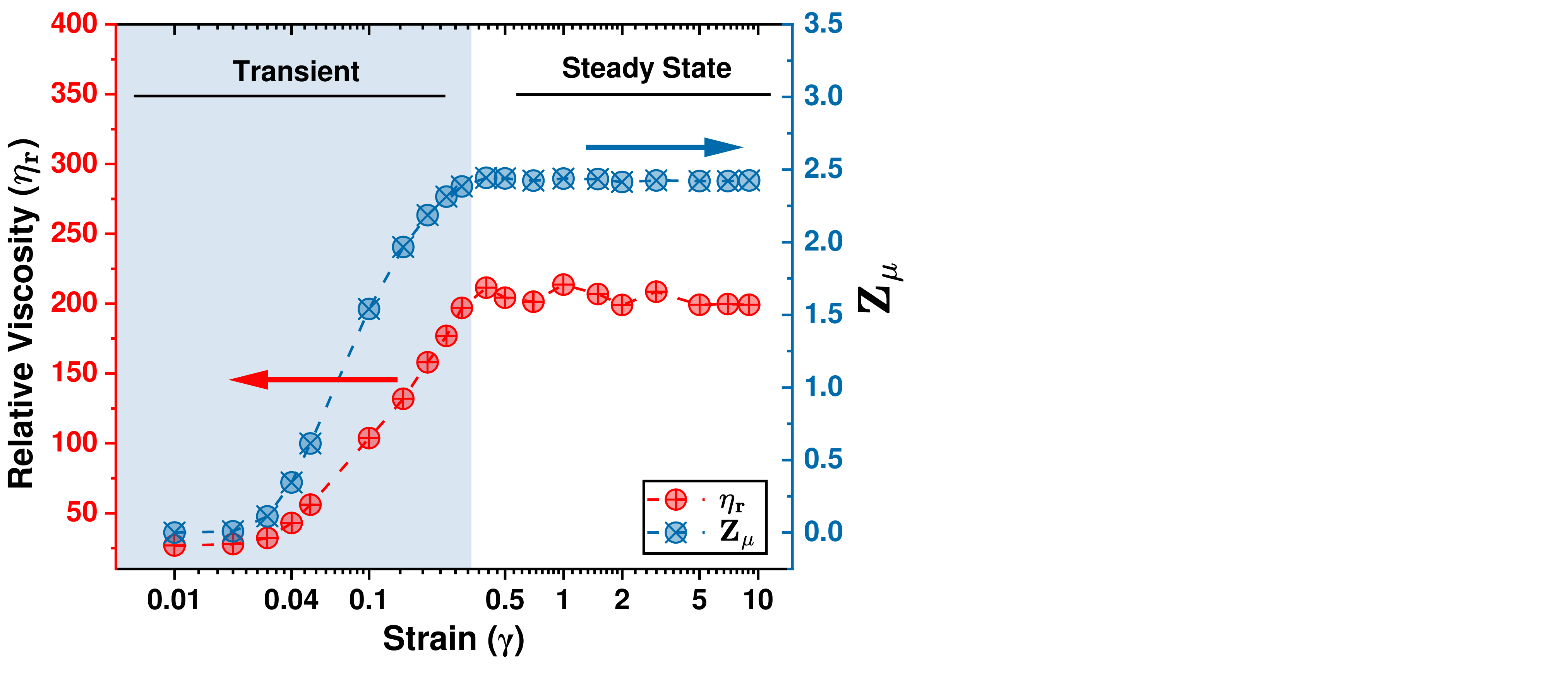}
    \caption{(a) Relative viscosity $\eta_r$ and average frictional coordination number $Z_\mu$ of suspensions as a function of strain $\gamma$. The suspension spends a short period of time in the transient regime ($\gamma \lesssim 0.35$) before reaching a dynamic steady-state $\gamma \gtrsim 0.45$. Here, the simulation is performed at $\phi$ = 0.76, and $\{\mu_s,\mu_r\}$ = \{0.5,0\}.}
    \label{fig:fig9}
\end{figure}

\begin{figure*}
    \centering
    \includegraphics[trim = 0mm 290mm 100mm 0mm, clip,width=0.95\textwidth,page=10]{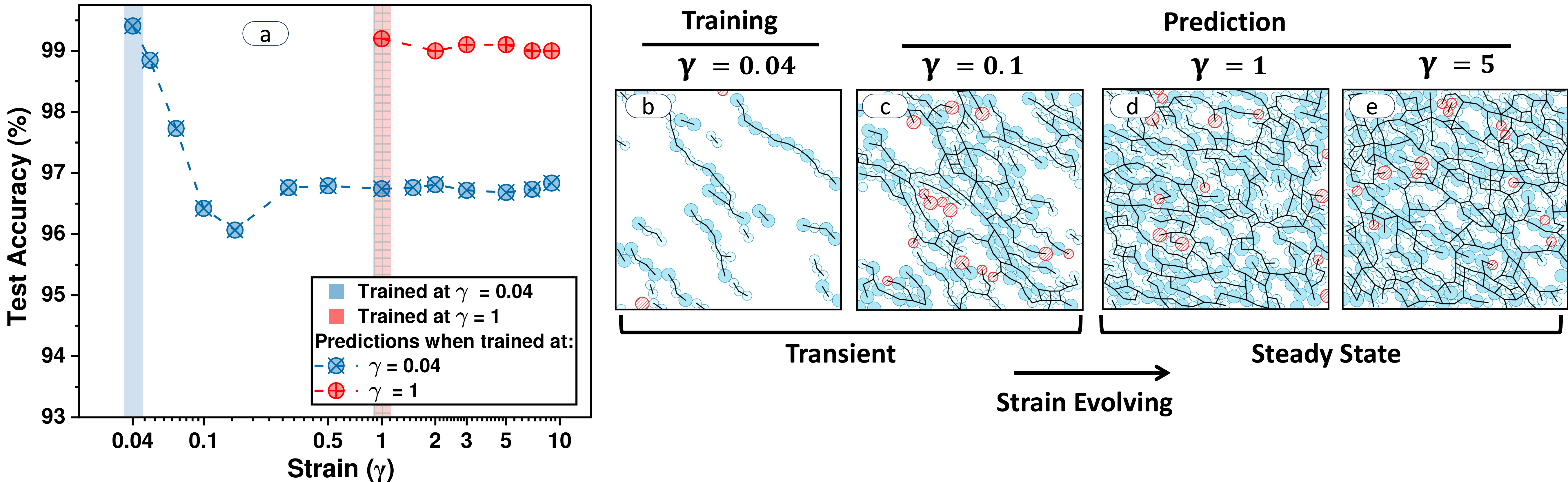}
    \caption{Prediction of frictional contact network (FCN) for future strains ($\gamma$) in suspension. (a) Test accuracy results for prediction of the FCN when the model is trained at transient, $\gamma$ = 0.04 (blue) or steady state, $\gamma$ = 1 (red). (b-e) Visualization of the predictions for the model trained at $\gamma$ = 0.04 with (b) validation for training condition, (c-e) tested for unseen configurations at $\gamma$ = 0.1, 1, and 5, respectively. While the model is trained at transient condition ($\gamma$ = 0.04), it demonstrated extrapolation of a wide range of strains, capturing FCN transient ($\gamma \lesssim 0.35$) and steady-state $\gamma \gtrsim 0.45$. Here, the simulations are performed at $\phi$ = 0.76, and $\{\mu_s,\mu_r\}$ = \{0.5,0\}.} 
    \label{fig:fig10}
\end{figure*}

\section{Results}
Recent studies have established that the rheology of dense frictional particle suspensions is governed by the packing fraction $\phi$, applied shear stress $\sigma$, and constraints on relative particle motion $(\mu_s, \mu_r)$~\cite{Morris_2020, Singh_2023, Ness_2016, Singh_2024}; thus making the suspension rheology a highly non-linear function of $\phi,\sigma,\mu_s,\mu_r$. 
The following section first presents the rheological and microstructural properties of suspensions obtained from the simulations under fixed shear stress, systematically varying the key parameters.
 The rheological response is represented using the steady-state relative viscosity $\eta_r$, while the microstructure is characterized by the mean frictional coordination number $Z_\mu$. Additionally, visualizations of the frictional contact network (FCN), highlighting only the non-rattlers, i.e., particles with at least one contact, are presented.
Following this, we demonstrate the robustness of the GNN model in accurately predicting the FCN for parameter combinations not included in its initial training set, showcasing its capability for generalization beyond the training conditions.

\paragraph*{Predictions for packing fraction $\phi$ and shear stress $\sigma$:}
Figure~\ref{fig:fig3} presents the rheological and microstructural properties of suspensions as a function of the packing fraction $\phi$ for various values of dimensionless shear stress $\sigma/\sigma_0$. 
Figures~\ref{fig:fig3}~a and b display the relative viscosity $\eta_r$ and frictional coordination number $Z_\mu$, respectively, for interparticle friction coefficients $\{\mu_s,\mu_r\} = \{0.5, 0\}$. 
As expected, both $\eta_r$ and $Z_\mu$ increase with $\phi$, with the dependence being strongly sensitive to the applied dimensionless stress $\sigma/\sigma_0$. 
Notably, at $\{\phi,\sigma/\sigma_0\}= \{0.8,100\}$, the suspension approaches a \textit{nearly} shear-jammed state, indicated by nearly diverging viscosity and $Z_\mu \approx 3$, which corresponds to the isostatic condition in 2D~\cite{Silbert_2010}. In contrast, suspensions with other combinations of $\{\phi,\sigma/\sigma_0\}$ are in the flowing state.
Now we raise an important question: can the GNN model accurately predict the FCN in the nearly jammed state ($\{\phi,\sigma/\sigma_0\}= \{0.8,100\}$) when solely trained on data from the flowing regime? For instance, can it generalize from lower packing fractions at the same stress level  ($\sigma/\sigma_0=100$), to accurately infer frictional contacts at higher packing fractions $\phi$, especially $\phi=0.8$?

Next, we demonstrate that such predictions are indeed possible. Figure~\ref{fig:fig4} presents the FCN prediction results for suspensions at different values of $\phi$, while maintaining a constant applied shear stress $\sigma/ \sigma_0$. 
Figures~\ref{fig:fig4}a and b display the prediction accuracies when the GNN model is trained under two different conditions: at fixed $\phi = 0.70$ and $\phi = 0.80$, respectively. The blue highlighted region in both figures indicates the $\phi$ value at which the model is trained. Different symbols represent the prediction results for a fixed value of $\sigma/\sigma_0$, as mentioned in the legend. For instance, the blue crosses in Fig.~\ref{fig:fig4} a correspond to the prediction results when the model is trained at fixed
$\phi$ = 0.7, $\sigma/\sigma_0$ = 100 and predicted the FCN for higher $\phi$ values up to 0.8, while keeping $\sigma/\sigma_0$ constant at 100. 
The model classifies the particles contributing to the formation of the FCN by determining whether they are in frictional contact. Accordingly, the reported accuracy is defined as the ratio of correctly classified particles to the total number of particles.
Figures.~\ref{fig:fig4}c-n provide visualizations of both trained and predicted frictional contact networks (FCNs). In each case, the left panels (Figs.~\ref{fig:fig4} c, g, k and e, j, m) display snapshots of the suspension systems, where all particles are shown with solid colors representing particle radii, and misclassified particles highlighted in red. The right panels depict the corresponding FCN structures (e.g., Figs.~\ref{fig:fig4}d to Figs.~\ref{fig:fig4}c). Here, the frictional contacts are represented by black lines, and only particles with at least one frictional contact (non-rattlers) are shown. The rows correspond to different applied shear stresses at which the GNN model was trained.
Figures~\ref{fig:fig4}c and d show validation results for a model trained at a fixed condition {$\phi,\sigma/\sigma_0$ = 0.7, 5}, tested on an unseen configuration at $\phi = 0.8$ at the same stress $\sigma/\sigma_0 = 5$, Figs.~\ref{fig:fig4}g–j and Figs.~\ref{fig:fig4}k–n correspond to training at $\sigma/\sigma_0 = 50$ and 100, respectively.
Remarkably, despite significant differences in rheological behavior, the GNN model \textcolor{black}{conditioned at} a relatively semi-dilute condition ($\phi = 0.70$ and $\sigma / \sigma_0$ = 100) predicts the FCN for near-jammed conditions ($\phi = 0.80$ and $\sigma / \sigma_0$ = 100) with over 97\% accuracy. Note the highly anisotropic network for the training condition ($\phi = 0.70$ and $\sigma / \sigma_0$ = 100, Fig.~\ref{fig:fig4}i) with frictional contacts mostly along the compressive direction, while the predicted network (($\phi = 0.80$ and $\sigma / \sigma_0$ = 100), Fig.~\ref{fig:fig4}n) is less anisotropic with percolating contacts along both compressive and tensile directions.
This trend is consistent across other models conditioned at a fixed $\sigma / \sigma_0$ = 5, 10, and 50. 
Interestingly, models trained at $\phi = 0.80$ for varying $\sigma/\sigma_0$ values can also extrapolate FCNs for semidilute conditions. However, a slight reduction in accuracy is observed for models trained at lower stresses ($\sigma/\sigma_0$ = 5 and 10) compared to higher stresses ($\sigma/\sigma_0$ = 50 and 100) (Fig.~\ref{fig:fig4}b). This  discrepancy likely arises from microstructural differences between $\phi = 0.70$ and $\phi = 0.80$: as $\phi$ approaches jamming, the system becomes increasingly dense, forming a percolating contact network in both directions with reduced anisotropy.

We have demonstrated that the GNN can successfully predict the frictional contact network, even when trained under free-flowing conditions with variations in applied stress and packing fraction. In the rest of this study, we focus on the $\{\phi,\mu_s,\mu_r\}$ parameter space, eliminating stress dependence by considering the limit $\sigma/\sigma_0 \to \infty$ (setting $F_0=0$). We tune the normal stiffness $k_n$ to ensure that the average overlap remains below 0.5\% of the particle radius $a$, thereby maintaining the system in the rigid limit~\cite{Singh_2015, Mari_2014}.

\paragraph*{Predictions for Sliding Friction $\mu_s$:}
Next, we investigate the prediction of the force chain network across various combinations of $\{\phi,\mu_s\}$.
Figures~\ref{fig:fig5}a and b depict the rheological response and the frictional coordination number of suspensions as a function of sliding friction $\mu_s$ for various packing fractions $\phi$. Consistent with the prior studies, we observe an increase in $\eta_r$  with $\mu_s$~\cite{Mari_2014, Singh_2018, Ness_2016, Singh_2020,sivadasan2019shear}. Here, the color code represents the low ($\mu_s \le 0.05$), intermediate ($0.05<\mu_s<2$) and high friction ($\mu_s>2$) regimes. Notably, $\eta_r$ remains largely insensitive to $\mu_s$ in the low- and high-friction limits, with the most pronounced increase occurring in the intermediate friction regime. Figure~\ref{fig:fig5}b presents the evolution of $Z_{\mu}$ with $\mu_s$ for various packing fractions $\phi$. The observed trend is expected, as friction enhances the load-bearing capacity of contacts, preventing buckling of the contact network under shear. Consequently, $\eta_r$ increases with $\mu_s$ for a fixed $\phi$, as seen in Fig.~\ref{fig:fig5}a. $\eta_r$ for packing fraction $\phi=0.8$ essentially diverges for $\mu_s\ge 0.5$, with $Z$ nearly equal to the isostatic condition $Z_\mathrm{iso}=3$~\cite{Silbert_2010}. This behavior aligns with findings from three-dimensional studies in both dense suspensions~\cite{sivadasan2019shear, radhakrishnan2019force} and dry granular systems~\cite{Singh_2013}.
Despite the complex rheological behavior, the GNN model exhibits remarkable robustness, successfully predicting the FCN even under conditions that extend well beyond its initial training regime.

Figure~\ref{fig:fig6}a illustrates the prediction accuracy of the FCN for unseen configurations, where the model is trained solely at a fixed packing fraction ($\phi = 0.70, 0.76, 0.80$) and a low friction coefficient ($\mu_s = 0.01$), and then tested across a range of $\mu_s$ values up to 10. 
Conversely, Fig.~\ref{fig:fig6}b presents the test precision for the GNN model when trained at a high friction coefficient ($\mu_s=10$) for different packing fractions. In both cases, the GNN model achieves remarkable accuracy, exceeding 98\%. Notably, since the suspension at $\phi$ = 0.80 and $\mu_s \geq 0.5$ are in a shear-jammed state, the model for $\phi=0.8$ was trained at $\mu_s=0.5$.
Figures~\ref{fig:fig6}~c-j provide visual representations of the FCN for both the training and prediction cases. Figures.~\ref{fig:fig6}c and d illustrate an example of a validation set where the model trained at a fixed $\{\phi,\mu_s\} = \{0.7, 0.01\}$. Meanwhile, Figs.~\ref{fig:fig6}e and f are visualizations for a tested configuration at the same $\phi = 0.70$ but for $\mu_s$ = 10.
Here, Fig.~\ref{fig:fig6}c shows the complete suspension system, including all particles, while the right panel (Fig.~\ref{fig:fig6}d) displays the FCN, highlighting only particles with at least one frictional contact.
Similarly, Figs.~\ref{fig:fig6}(g-j) show the same as (c-f) but trained at \{$\phi$ = 0.80, $\mu$ = 0.01\} and tested at $\mu_s$ = 0.5 (Figs.~\ref{fig:fig6}(i-j)). 
The results in Figs.~\ref{fig:fig6}a and b underscore the robustness of the GNN in the extrapolation of the FCN despite the substantial variations in the microstructure due to frictional effects. Remarkably, the model trained under the \textit{nearly} frictionless case is capable of accurately predicting FCNs even in highly frictional, \textit{ near} jammed states, and vice versa, demonstrating its strong generalization capability across a wide parameter space.

%
%

\paragraph*{Predictions for Rolling Friction $\mu_r$:}
Next, we examine the ability of our GNN model to predict the FCN for particles with rolling friction $\mu_r$, a crucial factor in modeling real-life suspensions with adhesive surface chemistries and/or ``rough'' particle shapes~\cite{Guy_2018, Singh_2020, Singh_2022, Singh_2022a}.
Figure~\ref{fig:fig7} presents the relative viscosity $\eta_r$ and the frictional coordination number $Z_\mu$ as a function of $\mu_r$ for $\{\phi,\mu_s\} = \{0.7, 0.5\}$. 
As shown, even small values of $\mu_r$ significantly impact $\eta_r$ and $z_\mu$, particularly for $\mu_r \geq 0.1$, emphasizing the strong dependence of suspension behavior on rolling friction.
Notably, for $\mu_r > 0.5$, the suspension reaches a shear-jammed state even at a relatively low packing fraction ($\phi=0.7$), highlighting the critical role of rolling friction. The frictional coordination number initially increases with rolling friction for $\mu_r<0.2$, similar to the behavior observed for $Z_\mu$ as a function of $\mu_s$ (Fig.~\ref{fig:fig5}). 
However, for a higher value of $\mu_r$, a slight decrease in $Z_\mu$ is observed. 
This could be attributed to the increase in the number of rattlers with rolling friction, and such a system could reach a jammed state with a significant fraction of non-load-bearing particles. Similar behavior has been reported by Santos \textit{et al.}~\cite{Santos_2020} in three-dimensional granular simulations with rolling friction.

Figure ~\ref{fig:fig8}a shows the FCN prediction accuracy across different values of $\mu_r$ with the GNN model trained at zero rolling friction (blue) or a small (but finite) rolling friction $\mu_r = 0.01$ (red).
As shown, the model can successfully extrapolate the FCN for $\mu_r$ values ranging from 0 to 0.5 when trained at $\mu_s = 0.5$, $\mu_r$ = 0 or 0.01. We do not examine cases with $\mu_r \geq 0.5$as the suspension enters a shear-jammed state beyond this threshold.
Figures~\ref{fig:fig8}b and c illustrate the validation results for the training condition ($\mu_r = 0$), while Figs.~\ref{fig:fig8}d-e and Figs.~\ref{fig:fig8}f-g show the model's extrapolation capabilities for $\mu_r = 0.01$, $\mu_r = 0.5$, respectively. 
%
%
For $\mu_r = 0$, the particles experience only sliding friction ($\mu_s = 0.5$). In contrast, the combination of $\{\mu_s,\mu_r\} = \{0.5, 0.5\}$ can be representative of a suspension composed of either faceted particles or those with extreme asperities~\cite{Singh_2020, Singh_2022}, leading to a highly anisotropic state approaching jammed. 

These findings demonstrated the ability of the model to generalize across physically distinct suspensions. It can be extended comprehensively by manipulating \{$\mu_s$, $\mu_r$\}, facilitating a rapid and efficient acquisition of information on the contact information of particles and FCN, and capturing various real-world scenarios where a multitude of grains with different types of friction coexist.

\paragraph*{{Prediction for Future Strains($\gamma$)}:}
Up to this point, all predictions have been based on training and prediction conditions within a \textit{dynamic} steady state.  Next, we explore a more intriguing question: Are predictions for a future condition possible? In other words, can we predict the steady-state condition based on transients? This is crucial for simulations since simulating the transients is less computationally expensive than simulating the suspension until the full steady-state.
Accurately capturing FCN at each time step in an evolving suspension system is essential for identifying transient behavior, ensuring physical consistency, and obtaining insights into the eventual steady state. 
Figure~\ref{fig:fig9} presents the relative viscosity $\eta_r$ and the average frictional coordination number $Z_\mu$ as a function of strain $\gamma$. As shown, the suspension remains in the transient regime up to approximately $\gamma \approx$ 0.35. Beyond $\gamma \gtrsim 0.45$, it reaches a fluctuating steady state, where both viscosity and coordination number are constant within their inherent fluctuations.

Figure~\ref{fig:fig10}a illustrates the test accuracy of extrapolation of the FCN at strains well beyond the original training conditions. Figures~\ref{fig:fig10}b-e provide visual representations of the predictions encompassing transient and steady-state with Fig.~\ref{fig:fig10}b, validation snapshot for the training condition ($\gamma = 0.04$), Figs.~\ref{fig:fig10}c to e, tested for $\gamma$ = 0.1, 1, and 5, respectively.
As shown, despite being trained only on initial time steps, $\gamma$ = 0.04, where the suspension remains in a transient regime, the model successfully predicts FCN for suspensions at subsequent strains, accurately capturing both transient and steady-state behaviors with a remarkable prediction accuracy exceeding 96\%. Moreover, when the model is trained at a steady state condition ($\gamma=1$), its predictive capabilities improve for future strains. The slight variations in accuracy in this scenario can be attributed to the inherent fluctuations observed in suspension under fixed shear stress, as shown in Fig.~\ref{fig:fig9}. 
These results underscore the efficiency of the model in capturing microstructure dynamics from the initial transients. Initially, the FCN is composed of linear friction contact chains, but as the suspension reaches the steady state, it forms loops, as illustrated in~\ref{fig:fig10}b-e. Moreover, the prediction results are directly influenced by the training conditions, as reflected in the evolving microstructure.
Note that training at very small strains ($\gamma<0.04$) is not physically feasible due to the lack of enough particles in frictional contacts in suspension (Figure~\ref{fig:fig9}). 
Overall, these findings demonstrate the robustness of the model in predicting the FCN at future strains. Given the high computational cost of resolving interparticle interactions during the transition to steady state, this model presents a promising approach for future applications, potentially reducing reliance on expensive parallel computing methods.


\section{Concluding Remarks}
This study introduced a DeepGNN-based model capable of accurately capturing the frictional contact network (FCN) structure in suspensions without explicit knowledge of interparticle forces. 
Although many data-driven algorithms struggle to extrapolate beyond training conditions, our approach leverages simulation data to overcome this limitation. The results demonstrate that the proposed DeepGCN model is a robust tool for inferring intricate interparticle interactions and underlying patterns, enabling the extrapolation of FCN to conditions far from the initial training settings. 
Remarkably, the model requires only particle radius and the relative distance of the particles as input, yet it generalizes to predict the FCNs across a wide range of conditions, including variations in surface roughness ($\mu_s,\mu_r$), strain ($\gamma$), and packing fraction ($\phi$) with various shear stresses ($\sigma_\text{xy}$), while each is trained at a single fixed value.
This generalization capability allows the model to capture the key parameters affecting rheological behavior and microstructural evolution, including transitions from transient to steady state. Furthermore, the model exhibits strong robustness across diverse conditions and parameter space, from idealized smooth particles to those with extreme roughness, spanning semi-dilute to near-jammed regimes.

Although this study focuses on a suspension system with spherical particles and a given particle size distribution, the proposed approach presents a powerful and generalizable framework for rapid and cost-effective predictions in a wide range of particulate systems. These include dry granular materials, complex suspensions such as particles with polymer brushes~\cite{bossis2022discontinuous}, dense fiber suspensions~\cite{khan2023constitutive}, and colloidal systems~\cite{smith2024topological, whitaker2019colloidal, nabizadeh2024network, nikoumanesh2023effect}.
Beyond suspensions, this method holds significant potential for predicting properties in complex engineering materials, including polymers to foster innovation for polymer discovery~\cite{qiu2024heat, gurnani2023polymer}, which can be further studied for the prediction of polymer blends and composites properties~\cite{majd2020curing, sohrabian2024molecular} where the bulk property emerges from intricate multi-scale interactions of their diverse components.
This versatile method can be extended to three-dimensional and real-world systems for optimization processes, real-time applications, and quantitative property predictions before engaging in costly simulation or experimentation. By bridging the gap between computational efficiency and physical accuracy, this approach opens new avenues for designing and understanding complex particulate systems.

\paragraph*{Acknowledgments}
All of this work made use of the High-Performance Computing Resource in the Core Facility for Advanced Research Computing at Case Western Reserve University.
A. S. acknowledges the Case Western Reserve University for start-up funding.

\bibliography{dst}

\end{document}